\documentclass{aa}

\usepackage{graphicx}
\usepackage{latexsym}
\usepackage{amssymb}
\usepackage{times}
\usepackage{natbib}
\bibpunct{(}{)}{;}{a}{}{,}

\begin{document}

\title{NGC 3310, a galaxy merger?}

\author{M. Kregel \inst{1}
   \and R. Sancisi \inst{1,2}}

\institute {Kapteyn Astronomical Institute, University of Groningen,
 P.O. Box 800, 9700 AV Groningen, The Netherlands
\and Osservatorio Astronomico, Bologna, Italy}

\offprints{M. Kregel}
\mail{kregel@astro.rug.nl}

\date{Received 19 March 2001/ Accepted 21 June 2001}

\abstract{The \ion{H}{i} structure and kinematics of the peculiar
starburst galaxy \object{NGC 3310} (\object{Arp 217}, \object{UGC
5786}) are discussed. New evidence bearing on the origin of the
starburst is presented. The bulk of \ion{H}{i} coincides with the
bright optical disk and shows differential rotation. Its velocity
dispersion is, however, unusually large for a spiral galaxy (up to
$\simeq$ 40 km s$^{-1}$), suggesting that the disk is highly perturbed
as already indicated by optical emission line spectroscopy. There are,
in addition, two prominent \ion{H}{i} tails, one extending to the
north-west and the other, somewhat patchy, to the south. These
\ion{H}{i} tails, the perturbed kinematics and the peculiar optical
morphology strongly suggest a recent merger between two gas-rich
galaxies. This seems to have been a major merger in which most of the
gas in the inner parts has been preserved in neutral atomic form and
either one of the progenitor disks has survived or a new disk has
formed.
}

\maketitle

\section{Introduction}
\begin{figure}
	\centering
	\resizebox{8.0cm}{!}{\includegraphics{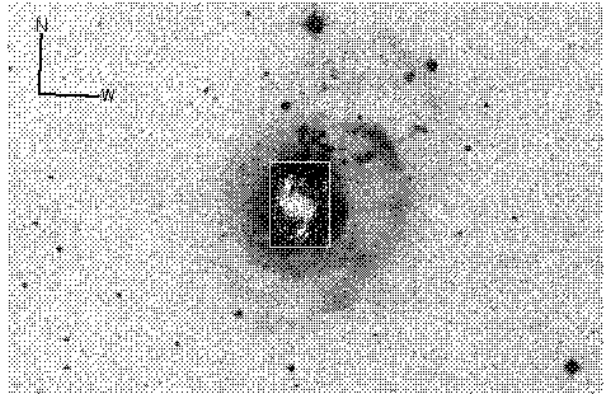}}
	\caption{An H$\alpha$ image (taken by H.C. Arp)	overlayed on a
	negative optical {\it B}--band image (both images are from van
	der Kruit and de Bruyn, 1976). The length of the N--W
	indicators is 1 arcmin (= 3.9 kpc).}
	\label{fig:overlay}
\end{figure}

The galaxy \object{NGC 3310}, located in the vicinity of the Ursa
Major cluster, lies at a distance of $\sim$13 Mpc\footnote{All distance
dependent parameters are calculated using $H_{0} = 75$ km s$^{-1}$
Mpc$^{-1}$} and is classified as SABbc(r)pec \citep{V91}. Van der Kruit
and de Bruyn \citeyearpar{vdKB76} have investigated its group membership and 
its environment and have concluded that it has no companions. Its main
global properties are listed in Table~\ref{tab:globalprop}. 
\object{NGC 3310} is a relatively small system undergoing a strong
starburst \citep{TG84,S96}. The optical morphology is illustrated in
Fig.~\ref{fig:overlay}. The bright inner region is dominated by a
two-armed open spiral pattern in H$\alpha$
\citep{vdKB76,BH81,MD96}. The inner part of this well-developed
pattern connects to a $\sim900\ \mbox{pc}$ diameter starburst ring,
surrounding the blue compact nucleus. The circumnuclear regions show a
moderately low metallicity, whereas the nucleus has solar abundances
\citep{P93}. The Far UV and {\it B}--band surface brightness profiles
of \object{NGC 3310} are very similar and seem to follow an {\it
R}$^{1/4}$ law outside the inner starburst ring \citep{S96}. The
outer parts of \object{NGC 3310} are dominated by the so-called
`bow-and-arrow' structure \citep{WC67}. The diffuse `bow' at  the
western side is a ripple (at $\sim$6 kpc from the center) consisting
primarily of late-type stars, possibly debris from an accreted disk
\citep{SS88}. A second fainter ripple can be seen farther out to the
north-west. The `arrow' (extending from 4 to 9 kpc from the center to
the north-west) is a chain of bright knots consisting of stellar
clusters most likely containing young massive stars and has a similar
age as the central starburst ring \citep{S96}. The `bow-and-arrow'
structure has been interpreted as a one-sided jet emanating from the
nucleus \citep{BS84} or as the result of the accretion of a small
gas-rich galaxy \citep{BH81,SS88,MDB95,S96}. Besides the `arrow',
other knots with recent star formation are seen at the northern and
southern part of the disk \citep[][ their fig.~1]{vdKB76}.

\begin{table}
	\caption{Properties of \object{NGC 3310}}
\begin{flushleft}
\begin{tabular}{llll}\hline\noalign{\smallskip}\hline\noalign{\smallskip}
Quantity        & Units         & \object{NGC 3310}      & Ref.$^{1}$\\
\hline\noalign{\smallskip}
Names		&		& UGC 5786	         & \\
		&		& ARP 217	         & \\
		&		& PGC 31650	         & \\
Hubble type	&               & SABbc(r)pec            & V91\\
$\alpha_{1950}$ & 	        & $10^{\rm h}35^{\rm m}40.3^{\rm s}$ & V91\\
$\delta_{1950}$ & 	        & $53^{\rm d}45^{\rm m}45^{\rm s}$   & V91\\
distance (adopted)& Mpc         & 13.3                   & \\
$D_{25}$	& arcmin	& $3.1\pm0.1$	         & V91\\
		& kpc           & $11.9\pm0.4$           & \\
$B_\mathrm{T}^{0}$        & mag                          & 10.92      & V91\\
$L_\mathrm{B}$$^{2}$      & $10^{10}\ {\rm L}_\mathrm{B,\sun}$ & 1.18       & \\
$(B-V)_\mathrm{T}^{0}$    & mag                          & 0.32	      & V91\\
$L_{\rm FIR}$             & $10^{10}\ {\rm L}_\mathrm{\sun}$ & 1.10       & S96\\
SFR$_\mathrm{global}$     & M$_\mathrm{\sun}$ yr$^{-1}$  & 8.5        & S96\\
inclination angle (H$\alpha$) & degrees                  & $32\pm6$   & K76\\
inclination angle (\ion{H}{i})& degrees                  & $52\pm2$   & M95\\
position angle (H$\alpha$)    & degrees                  & $172\pm4$  & K76\\
position angle (\ion{H}{i})   & degrees                  & $163\pm3$  & M95\\
\hline\noalign{\smallskip}\hline\noalign{\smallskip}
\end{tabular}

\begin{enumerate}
\item Acronyms; K76 -- \citet{vdK76}, V91 -- \citet{V91}, M95 --
\citet{MDB95} and S96 -- \citet{S96}. \item The {\it B}--band
magnitude is converted to luminosity in solar units using
$\mbox{M}_\mathrm{B,\sun} = 5.48$.
\end{enumerate}
\end{flushleft}
	\label{tab:globalprop}
\end{table}

The ionized gas shows large streaming motions (up to 60 km s$^{-1}$)
along the arms \citep{vdK76,GS91}, which are consistent with the
presence of a strong density wave \citep{vdK76,MD96}. The H$\alpha$
rotation curve in the nuclear region is one of the steepest rising
rotation curves found in spiral galaxies \citep{vdK76,GS91}. Another
peculiarity is the offset between the dynamical center and the stellar
nucleus of $(96\pm19)$ pc \citep{WC67,vdK76,BH81}.

Radio continuum maps \citep{vdKB76,BH81,D86} show extended, remarkably
bright synchotron emission from the inner regions and strong sources
coincident with the giant \ion{H}{ii} regions in the arms, in the
inner ring, and in the nucleus. The strong IR emission correlates well 
with both the continuum and H$\alpha$ emission \citep{TG84}.

An earlier study of the neutral hydrogen \citep{MDB95} revealed an
\ion{H}{i} extension of the optical `arrow' out to $\sim24$ kpc and
extended \ion{H}{i} at $\sim37$ kpc to the south of the nucleus with
no optical counterpart. Another peculiarity was a `hole' in the
\ion{H}{i} distribution centered on the south-eastern part of the
optical ring (SN 1991N is situated near the center of this
`hole'). The molecular gas shows a clumpy distribution and is
primarily associated with the spiral arms \citep{K93,MDB95}. In the 
nucleus it is less clear: a small amount of molecular gas is not ruled
out. The total H$_{2}$ mass, obtained using the Galactic CO--H$_{2}$
conversion factor, is $\sim$ 2 10$^{8}$ M$_\mathrm{\sun}$
\citep{K93,MDB95}, a value typical for late-type galaxies.

In summary, many features of \object{NGC 3310} -- the unusual
morphology, the starburst features, the streaming of ionized gas along
the arms, the offset between the dynamical center and the nucleus, the 
ripples in the outer parts, the `arrow' and its \ion{H}{i} counterpart
-- indicate that some major disturbance has affected gas and stars and
has led to massive star formation. These features together with the
Far UV and {\it B}--band {\it R}$^{1/4}$ surface brightness profiles
indicate that \object{NGC 3310} may well be the result of a merger
event.

New \ion{H}{i} data, obtained with the Westerbork Synthesis Radio
Telescope (WSRT) and combined with those of \citet{MDB95} are
presented here. They give additional information concerning the origin
of the starburst and of the optical and \ion{H}{i} peculiarities.

\begin{table}[!t]
	\caption{WSRT Observing Parameters}
\begin{flushleft}
\begin{tabular}{lll}\hline\noalign{\smallskip}\hline\noalign{\smallskip}
Observation 			& A 		& \\
\hspace{2cm} 12hrs, 36m 	&		& 16 July 1997\\
\hspace{2cm} 12hrs, 72m 	&		& 11 June 1997\\
        	                & B 		& \\
\hspace{2cm} 12hrs, 72m 	&		& 15 Jan. 1987\\
Field centers 		        & A 		& \\
\hspace{2cm} $\alpha_{1950}$    & 		& $10^\mathrm{h}35^\mathrm{m}40.00^\mathrm{s}$\\
\hspace{2cm} $\delta_{1950}$    & 		& $53^\mathrm{d}43^\mathrm{m}0.01^\mathrm{s}$\\
        	                & B             & \\
\hspace{2cm} $\alpha_{1950}$    & 		& $10^\mathrm{h}35^\mathrm{m}40.1^\mathrm{s}$\\
\hspace{2cm} $\delta_{1950}$    & 		& $53^\mathrm{d}45^\mathrm{m}49^\mathrm{s}$\\
Central velocity (km s$^{-1}$)  & A             & 970.00\\
			        & B             & 1000.00\\
Baselines (m)	                & A 		& 36:2736:36\\
			        & B 		& 36:2700:72\\
Bandwidth (MHz)		        & 		& 5\\
Number of channels	        &		& 63\\
Channel separation (km s$^{-1}$)&       	& 16.6\\
Velocity weighting              & A		& Uniform\\
			        & B		& Hanning\\
Synthesized beam (FWHM)         & C 		& $14\arcsec.2\times17\arcsec.7$\\
Velocity resolution (km s$^{-1}$) (FWHM) & C          & 33.3\\
Noise level (1$\sigma$) (mJy beam$^{-1}$) & C		& 0.8\\
\hspace{2.14cm}(K)		&		& 1.92\\
\hline\noalign{\smallskip}\hline\noalign{\smallskip}
\end{tabular}
\end{flushleft}
	\label{tab:obsparameters}
\end{table}

\section{Observations and Data Reduction}
The new 21-cm line observations of \object{NGC 3310} were obtained
with the WSRT as a part of WHISP (Westerbork \ion{H}{i} Survey of
Spiral and Irregular Galaxies) \citep{SASH01}. Due to telescope
maintainance only 27 interferometers were available during these
observations.

The main observational parameters are listed in
Table~\ref{tab:obsparameters}, where `A' denotes the WHISP
observations, `B' the observations obtained by \citet{MDB95} and `C'
the combination of `A' and `B' (described below). The combined
observations were smoothed to $29\arcsec.9\times30\arcsec.7$ and
$60\arcsec.9\times63\arcsec.0$ to study the extended, low surface
brightness structures.

\indent
The map of the radio continuum was obtained from a linear fit of a
baseline over the line free channels on either side of the band (28
channels in total) and was subsequently subtracted from the
datacube. The resulting line maps were cleaned to correct for
instrumental effects like sidelobes, grating rings and offset
baselevels. The search area was determined for each channel map
separately, in some cases after smoothing to 90\arcsec\ to identify
the regions containing extended emission. After cleaning, the new data
and those of \citet{MDB95} were combined.

The \ion{H}{i} channel maps are shown in Fig.~\ref{fig:channels}. The
contours show the distribution of \ion{H}{i} intensity at
30\arcsec\ resolution. The extended emission is shown at 60\arcsec\
resolution by the shaded areas. The optical picture and the `dirty'
30\arcsec\ radio continuum map are also shown.

\section{Analysis}

\subsection{Radio Continuum}
\begin{figure}[!t]
	\resizebox{7cm}{!}{\includegraphics{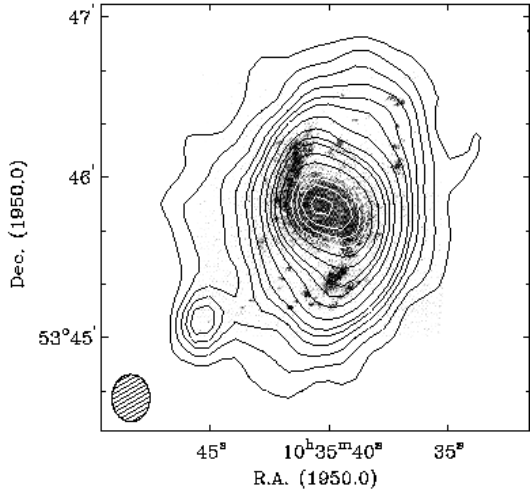}}
	\caption{Full resolution ($14\arcsec.2\times17\arcsec.7$) map of the
	radio continuum emission from \object{NGC 3310} at 1.42
	GHz. The r.m.s. noise is $0.29\ \mbox{mJy}\ \mbox{beam}^{-1}$, contour
	levels are (first:last:increment) 1:6:1, 8:17:3, 23:51:7 and 59
	$\mbox{mJy}\ \mbox{beam}^{-1}$. The H$\alpha$ image (van der
	Kruit and de Bruyn, 1976) is shown in greyscale.}
	\label{fig:contin}
\end{figure}

\noindent
Figure~\ref{fig:contin} shows the cleaned radio continuum map of
\object{NGC 3310} at full resolution. The extended source of emission
includes nucleus, central ring and spiral arms; its peak surface
brightness is 63 mJy beam$^{-1}$. The 5.5 mJy beam$^{-1}$ point
source, visible $\sim$1\arcmin\ south-east of the nucleus, is probably
a background source. The position of the nucleus cannot be determined
accurately for such an extended source. The value of the continuum
flux at 1.42 GHz (Table~\ref{tab:measprop}) agrees with those
published by \citet{L71} and by \citet{D86}. The total power is
$21.85\pm0.03$ ($\log$ (P$_{1.42 \mathrm{GHz}}$/ W Hz$^{-1}$)).

\begin{figure}[!t]
	\resizebox{7.5cm}{!}{\includegraphics{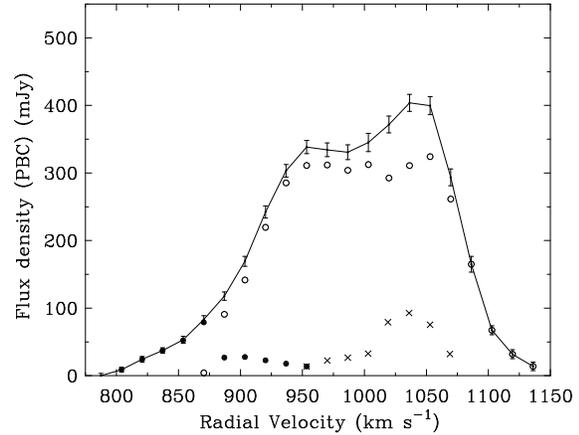}}
 	\caption{Global 21cm line profile of \object{NGC 3310} (solid
 	line). The dots show the northern tail, the crosses
 	the southern tail and the circles the \ion{H}{i} disk.}
	\label{fig:global}
\end{figure}

\begin{table}[!t]
	\caption{Integral \ion{H}{i} properties and the radio continuum flux of
	\object{NGC 3310}}

\begin{flushleft}

\begin{tabular}{llll}\hline\noalign{\smallskip}\hline\noalign{\smallskip}
\multicolumn{2}{l}{Quantity} & \multicolumn{1}{l}{Units} & \multicolumn{1}{l}{Value} \\ \hline\noalign{\smallskip}
$\int S\ dv$	                & & Jy km s$^{-1}$                  & $69\pm4$\\
$M_\mathrm{\ion{H}{i}}$	& & $10^{9}\ \mbox{M}_\mathrm{\sun}$& $2.8\pm0.2$\\
$S_{1.42\ \mbox{GHz}}$	& & Jy			            & $0.34\pm0.01$\\
\hline\noalign{\smallskip}\hline\noalign{\smallskip}
\end{tabular}
\end{flushleft}
	\label{tab:measprop}
\end{table}

\begin{figure*}[!ht]
	\centering
	\resizebox{17cm}{!}{\includegraphics{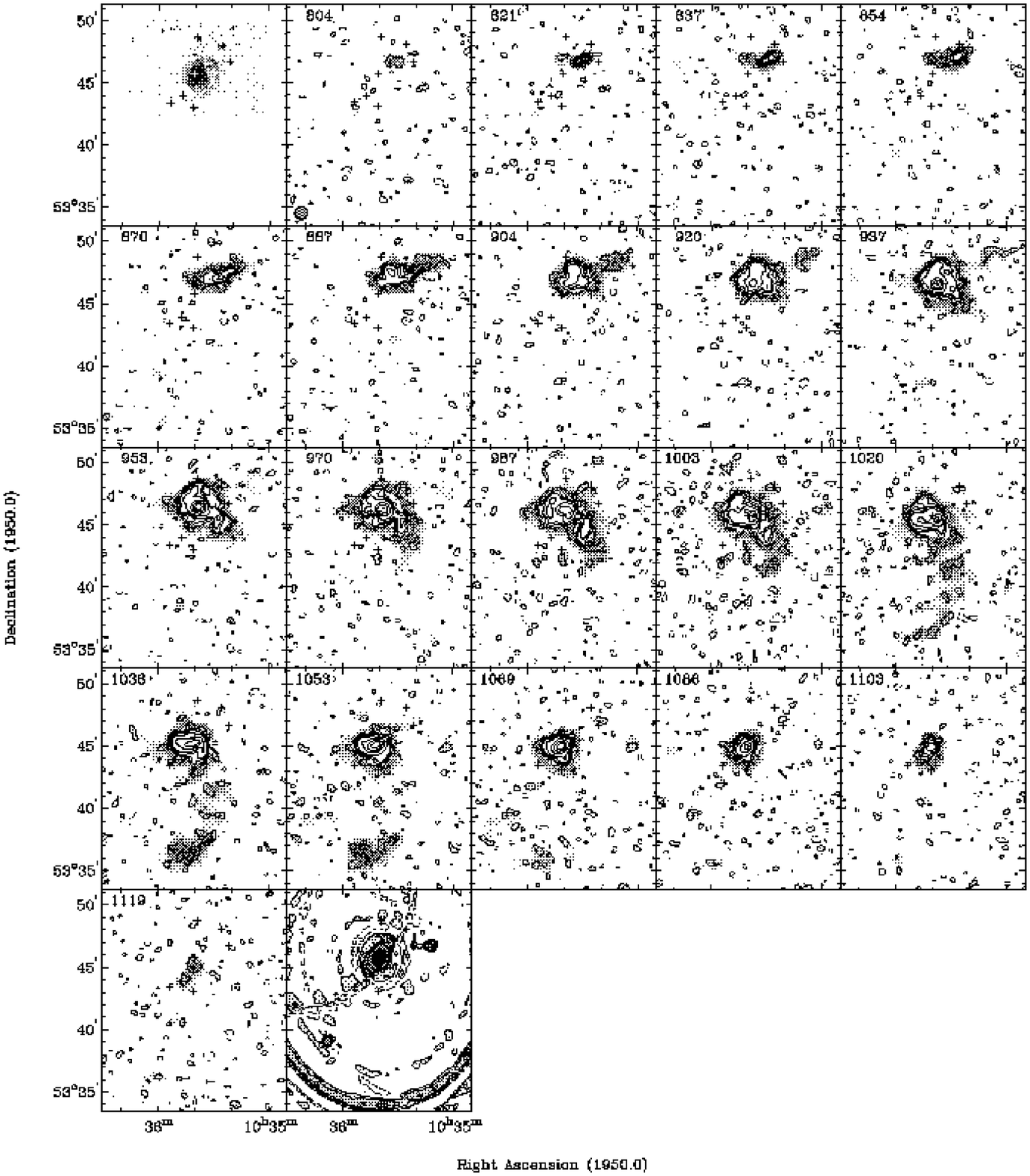}}
	\caption{\ion{H}{i} channel maps of \object{NGC 3310}. The
	beam size (HPBW) is $29\arcsec.9\times30\arcsec.7$. The
	contour levels are $-$3.2, $-$1.6, 1.6 ($\sim$2$\times$
	r.m.s. noise), 3.2, 4.8, 6.3, 7.8, 15.7, 23.6 and 31.4 mJy
	beam$^{-1}$. The shading shows the faint extended emission in
	the outer parts on a logarithmic scale up to 6.5 mJy
	beam$^{-1}$ at a resolution of
	$60\arcsec.9\times63\arcsec.0$. Also the optical image (top
	left) and the radio continuum image used for subtraction
	(bottom right) are shown. The radial velocities (in km
	s$^{-1}$) are heliocentric.}
	\label{fig:channels}
\end{figure*}

\subsection{Anomalous \ion{H}{i}}
\noindent
The \ion{H}{i} line flux was determined in each channel map at
60\arcsec\ resolution. The resulting global \ion{H}{i} profile is shown in
Fig.~\ref{fig:global}. It is similar to that obtained by
\citet{MDB95}. The asymmetric structure and the remarkable tail on the
low velocity side clearly point at some anomaly in the \ion{H}{i}
distribution and/or kinematics. The integral quantities derived from
the global profile are listed in Table~\ref{tab:measprop}.

The channel maps (Fig.~\ref{fig:channels}) show that the bulk of
\ion{H}{i} coincides with the bright optical disk and has the
characteristic pattern of differential rotation. There are, in
addition, two extended features, one on the north-west side visible in
the velocity range 870 to 953 km s$^{-1}$ and the other to the south,
clearly visible (shaded) at velocities 1003 to 1069 km s$^{-1}$. Both
have already been noticed and reported by \citet{MDB95}. The better
sensitivity of the data presented here makes it possible, however, to
significantly improve the study of their structure and kinematics.
In the following, these two extended features will be referred to
as the northern and the southern tail.

\vspace{0.2cm}
{\it a) Northern tail}
\vspace{0.2cm}

\begin{figure}[!ht]
	\resizebox{8cm}{!}{\includegraphics{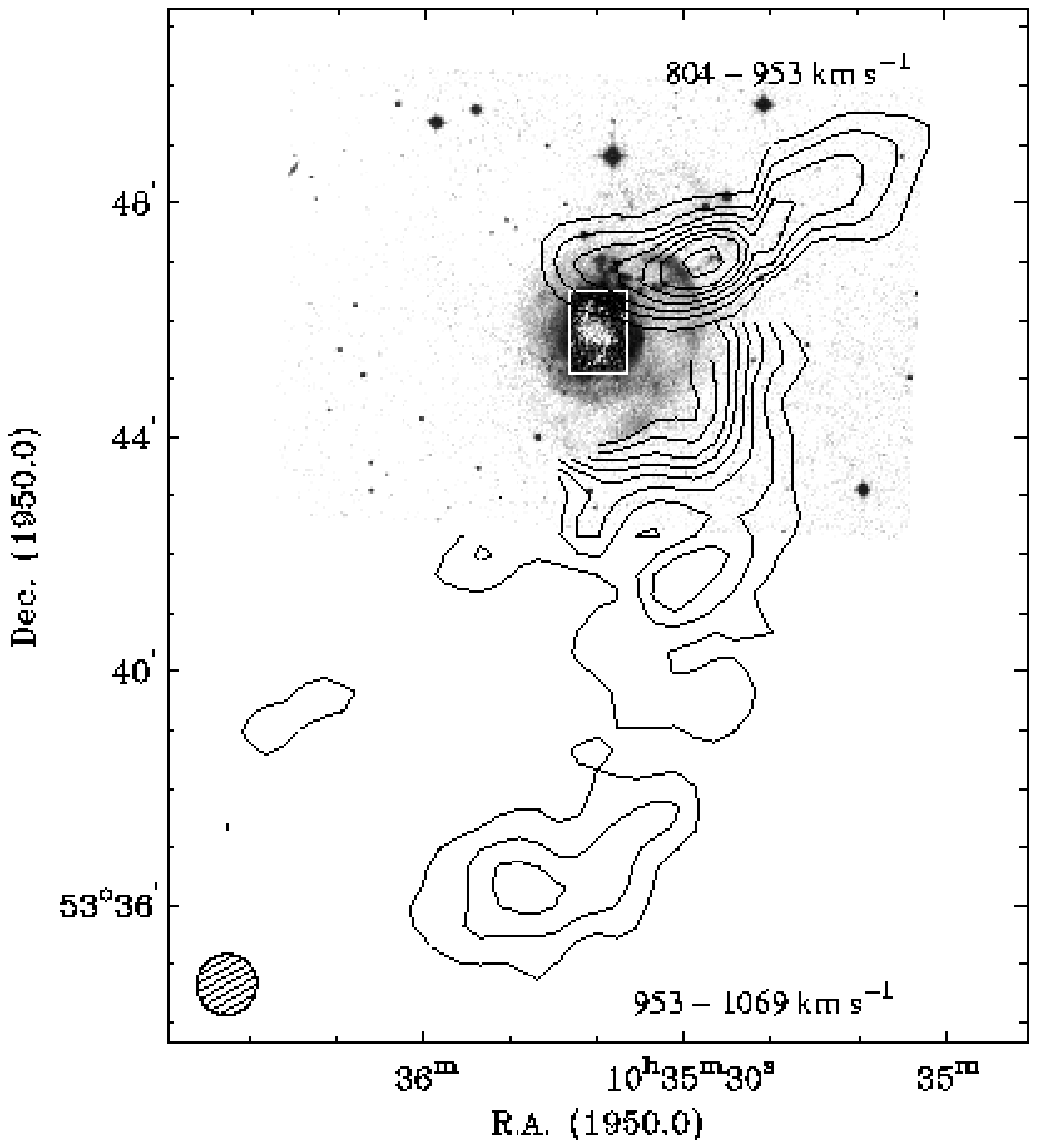}}
	\caption{Map showing the \ion{H}{i} tails of \object{NGC 3310}
	overlayed on the optical picture. The beam size is
	60\arcsec.9$\times$63\arcsec.0. The contour levels are 0.4, 0.8,
	1.2, 1.6, 2.0, 2.4, 3.0 and 3.6 $\times$10$^{20}$ cm$^{-2}$.
	The velocity ranges for the northern and the southern tail
	are indicated.}
	\label{fig:anomal}
\end{figure}

\begin{figure}
	\resizebox{7cm}{!}{\includegraphics{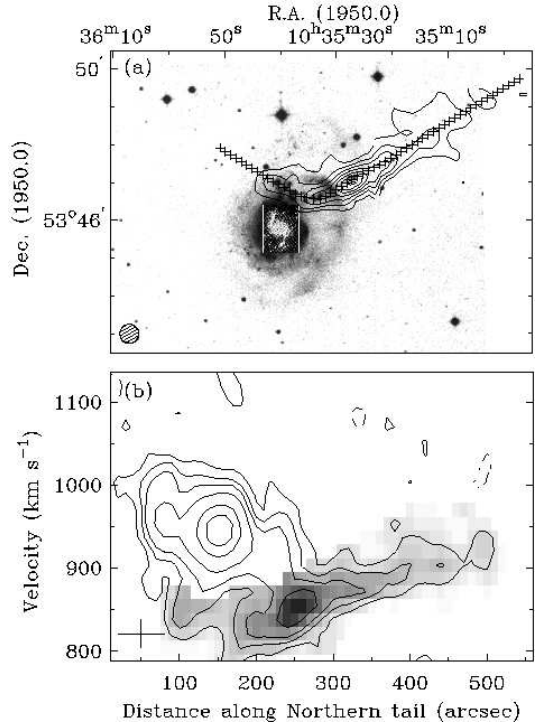}}
	\caption{(a) 30\arcsec\ contourmap of the northern tail
	overlayed on the optical picture. The contour levels are 1, 2,
	3, 4, 5, 6 and 7 10$^{20}$ cm$^{-2}$, (b) position--velocity
	map along the track from east to west marked in (a)
	following the ridge of the northern tail. The contour levels
	are as for the channel maps (Fig.~\ref{fig:channels}).
	In this position-velocity diagram the tail is shown shaded.}
	\label{fig:NWcuts}
\end{figure}

The \ion{H}{i} extension to the north-west (Fig.~\ref{fig:channels})
coincides with the already mentioned and well-known `arrow'. One
cannot fail to notice the continuity in space and velocity of this
feature with the emission in the channel maps at lower velocities,
from 854 to 804 km s$^{-1}$. It 
should also be noted that these are precisely the velocities of the
anomalous tail in the global \ion{H}{i} profile. It is therefore
natural to associate the \ion{H}{i} in the anomalous 804--854 km
s$^{-1}$ velocity range with the anomalous \ion{H}{i} pointing away
from the main body (the `arrow') in the adjacent, higher velocity
channels 870 to 953 km s$^{-1}$. The resulting structure has the shape
of a tail (see Fig.~\ref{fig:anomal}) located on the northern edge of the 
disk, slightly curved and elongated from the east to the
north-west. It has been obtained by adding the \ion{H}{i} over the
whole velocity range 804 to 953 km s$^{-1}$ and by taking, at velocities
887 to 953 km s$^{-1}$, only the emission from the arrow. There is a
striking coincidence of this \ion{H}{i} tail not only with the
optically bright knots of the `arrow' itself but also with the bright
compact knots seen along a curve on the northern side of the bright disk
of \object{NGC 3310}. Since the tail is unresolved at 60\arcsec, the same
procedure as described above was repeated at 30\arcsec\ and the result
is shown in Fig.~\ref{fig:NWcuts}a. The coincidence with the optical knots
is even clearer, suggesting a physical association of the knots with 
the \ion{H}{i} tail. These knots should then have the same velocity as
the \ion{H}{i}; but according to the H$\alpha$ data published by
\citet{MD96}, their velocities seem to be around 950 km s$^{-1}$,
difficult to reconcile with the velocity of the \ion{H}{i} indicated
here.

The kinematics and structure of the tail are illustrated in the
position--velocity map of Fig.~\ref{fig:NWcuts}b which was constructed
along the path marked by the crosses in Fig.~\ref{fig:NWcuts}a. The
distance along the tail in Fig.~\ref{fig:NWcuts}b is measured with
respect to the most eastern marker. The figure shows the \ion{H}{i} of
the disk, which is centered at 950 km s$^{-1}$, and the tail (shaded),
which is in the velocity range 800 to 940 km s$^{-1}$. There is a dip
in the tail at a distance of 150\arcsec\ and a velocity of 850 km
s$^{-1}$. The main properties of the tail are listed in
Table~\ref{tab:tailprop}. Its \ion{H}{i} mass is about 8\% of the
total hydrogen mass of \object{NGC 3310}.

Clearly, if the above analysis is valid and it is correct to connect the
kinematically anomalous \ion{H}{i} with that of the `arrow' as we have done,
a completely new physical picture emerges: instead of a straight jet
pointing outward from the nucleus as suggested before \citep{BS84}, we
have a curved tail-like structure attached to the northern edge of the
bright optical disk.\\

{\it b) Southern tail}
\vspace{0.2cm}

The channel maps in Fig.~\ref{fig:channels} show an extended, peculiar
feature (shaded area) on the southern side of \object{NGC 3310} at radial
velocities ranging from 953 up to $1069\ \mbox{km}\ \mbox{s}^{-1}$. It
has a somewhat patchy, but coherent tail-like structure
(see also Fig.~\ref{fig:anomal}) connecting smoothly to the \ion{H}{i}
disk on its western side. The southern structure shown in
Fig.~\ref{fig:anomal} was obtained by adding all channel maps in the
above velocity range.

\begin{figure}
	\resizebox{\hsize}{!}{\includegraphics{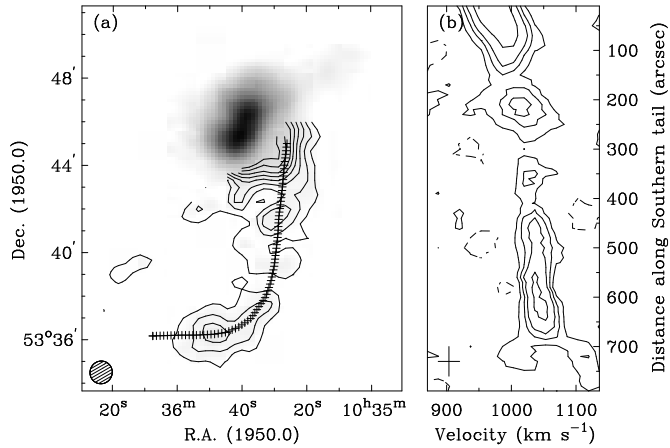}}
	\caption{(a) The southern \ion{H}{i} tail is
	represented in contours (same as in
	Fig.~\ref{fig:anomal}). The shading shows the total \ion{H}{i}
	density distribution of \object{NGC 3310} at 60\arcsec\ 
	resolution (obtained by integrating over all velocities),
	(b) position--velocity map along the track following the ridge
	of the tail from north to south, as marked in (a).
	The contour levels are $-$2.1, 2.1, 4.3, 6.4, and
	10.7 mJy beam$^{-1}$.}
	\label{fig:Scuts}
\end{figure}

Fig.~\ref{fig:Scuts}a shows the tail and the total \ion{H}{i}
at 60\arcsec\ resolution. The crosses mark the path along which the
position--velocity map shown in Fig.~\ref{fig:Scuts}b was
constructed. The distance along the tail in Fig.~\ref{fig:Scuts}b is
with respect to the most northern marker. The figure shows a clear
continuity in both space and velocity from the inner to the outer
regions. The tail seems to be an extension of the western part of the 
\ion{H}{i} disk (the emission closer to the bright disk has been
omitted because of confusion). Its main properties are given in
Table~\ref{tab:tailprop}. Its \ion{H}{i} mass is at least 9\% of the
total hydrogen mass of \object{NGC 3310}.

\begin{table}[!h]
	\caption{\ion{H}{i} properties of the tails of \object{NGC 3310}}
\begin{flushleft}
\begin{tabular}{llll}\hline\noalign{\smallskip}\hline\noalign{\smallskip}
Quantity & Units & Northern & Southern \\
\hline\noalign{\smallskip}
$M_\mathrm{\ion{H}{i}}$   & 10$^{8}$ M$_\mathrm{\sun}$ & $2.3\pm 0.2$ & $2.7\pm 0.2$ \\
Length                    & kpc                        & 23           & 51 \\
$\left(N_\mathrm{\ion{H}{i}}\right)_\mathrm{\mbox{max}}$ & 10$^{20}$ cm$^{-2}$ & 7.5 & $\sim$1.5 \\
$\sigma_\mathrm{\ion{H}{i}}$ & km s$^{-1}$ & 17 & 13\\
\hline\noalign{\smallskip}\hline\noalign{\smallskip}
\end{tabular}
\end{flushleft}
	\label{tab:tailprop}
\end{table}

In addition to the two well-developed \ion{H}{i} tails,
Fig.~\ref{fig:anomal} also shows hints of an isolated feature to the
south-east of \object{NGC 3310}, visible (faint shaded areas) in
Fig.~\ref{fig:channels} at velocities of 1053 and 1069 km
s$^{-1}$. This could be part of the southern tail described above. The
present observations, however, are not sensitive enough to be sure of
the reality of this feature.

\subsection{\ion{H}{i} disk}

The overall \ion{H}{i} density distribution and the velocity field in and
around \object{NGC 3310} are shown in Figs.~\ref{fig:moments}a and b.
More detailed maps of the \ion{H}{i} disk, at 20 arcsec resolution, 
are presented in Figs.~\ref{fig:moments}c and d. 
The high-resolution total \ion{H}{i} map was obtained
by defining the area of the emission for each channel map using the
lower-resolution maps as masks. The velocity field was obtained
by taking the density weighted mean with a 1.7 mJy beam$^{-1}$
(2$\sigma$) cutoff. In the construction of these maps, the emission from the
northern and southern tails, clearly visible in the
60\arcsec\  channel maps, was masked out in order to have a clearer picture
of the disk structure and kinematics.

In the 60\arcsec\ map, the central surface density reaches
$\sim$2.4 10$^{21}$ cm$^{-2}$ which corresponds to $\sim$11
M$_\mathrm{\sun}$ pc$^{-2}$ in the plane of the galaxy. This is
similar to the values found for normal high surface brightness
galaxies \citep{BW94,RA96}. The 20\arcsec\ resolution map shows that
the peak surface density is actually $\sim$4.5 10$^{21}$ cm$^{-2}$ and
is reached about 30\arcsec\ south of the optical center.

\begin{figure}[!t]
	\resizebox{8cm}{!}{\includegraphics{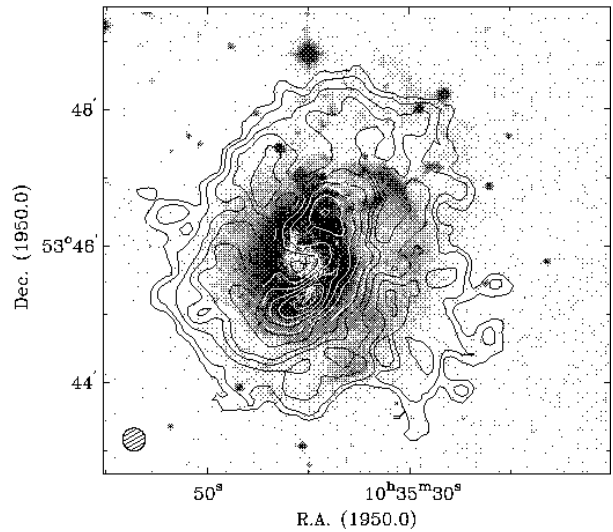}}
	\caption{20\arcsec \ion{H}{i} map (HPBW =
	$20\arcsec.0\times20\arcsec.0$) overlayed on the optical
	picture (Fig.\ref{fig:overlay}). The contour levels
	(first:last:increment, in units of 10$^{20}$ cm$^{-2}$) are
	1.5, 3:12:3, 18:42:6 and 45 as in Fig.~\ref{fig:moments}c.}
	\label{fig:20over}
\end{figure}

\begin{figure*}[!t]
	\centering
	\resizebox{15cm}{!}{\includegraphics{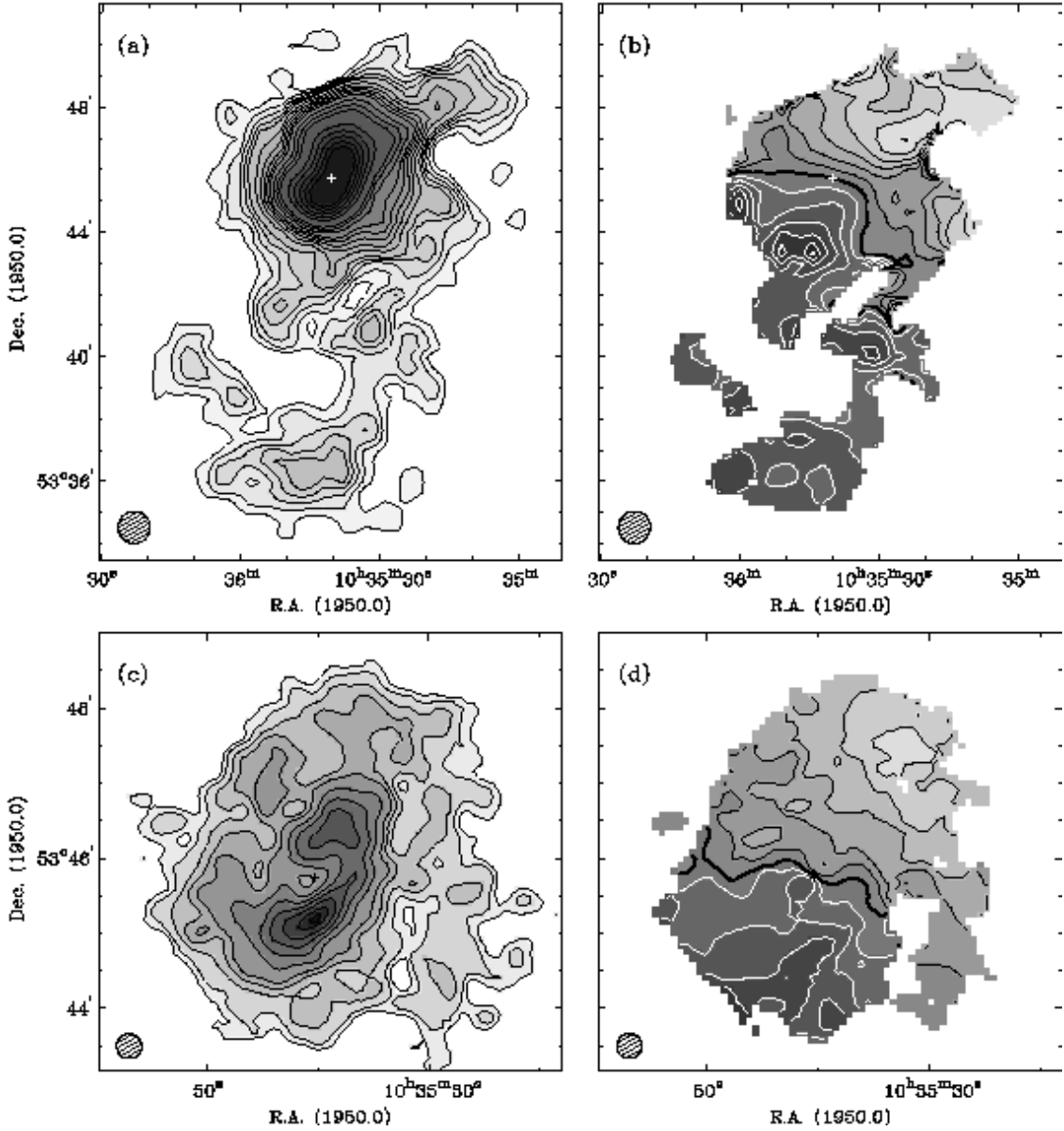}}
	\caption{(a) total \ion{H}{i} map (HPBW =
	$60\arcsec.9\times63\arcsec.0$). The contour levels
	(first:last:increment, in units of 10$^{20}$ cm$^{-2}$) are
	0.2:1.4:0.3, 2.0:4.4:0.6, 5.6:9.2:1.2 and 11.6:23.6:2.4, (b)
	velocity field (HPBW = $60\arcsec.9\times63\arcsec.0$). The
	contour levels are in steps of 20 km s$^{-1}$ and the thick
	contour represents v$_{\rm sys}$ at 1006 km s$^{-1}$ (the northern
	side is approaching) (c) \ion{H}{i} map (HPBW =
	$20\arcsec.0\times20\arcsec.0$), disk only. The contour levels
	are (in units of 10$^{20}$ cm$^{-2}$) 1.5, 3:12:3, 18:42:6 and
	45 (d) velocity field (HPBW =
	$20\arcsec.0\times20\arcsec.0$). The contour levels are in
	steps of 20 km s$^{-1}$ and the thick contour represents
	v$_\mathrm{sys}$ at 1006 km s$^{-1}$. The crosses in the
	panels mark the position of the optical center. The velocity
	fields are only defined where the surface densities are larger
	than the values of the second contour in the total \ion{H}{i}
	maps.}
	\label{fig:moments}
\end{figure*}

The 20\arcsec\ map (Fig.~\ref{fig:moments}c) shows a disk with a
bright inner part and a large depression centered $\sim$8\arcsec\
($\sim$ 500 pc) south-east of the optical center. This depression
coincides with the brightest region of Far UV emission \citep{S96}.
The bright inner part is surrounded on all sides by low surface density
structures (average densities about 5 M$_\mathrm{\sun}$ pc$^{-2}$).
In Fig.~\ref{fig:20over} this \ion{H}{i} map is compared with
the optical image. The highest \ion{H}{i} density features appear to
coincide with the extensions of the optical spiral arms (compare
Fig.~\ref{fig:20over} and Fig.~\ref{fig:moments}c). The outer low
surface brightness \ion{H}{i} seems to coincide only partly with the
optical ripple (the `bow') in the north-west and on the southern side.

The motion of the \ion{H}{i} disk is clearly dominated by
differential rotation although there are irregularities, especially in
the outer parts (Fig.~\ref{fig:moments}b and d). At 60\arcsec\
resolution such irregularities are partly due to the presence of the
anomalous \ion{H}{i}.

\begin{figure}[!t]
	\resizebox{7.5cm}{!}{\includegraphics{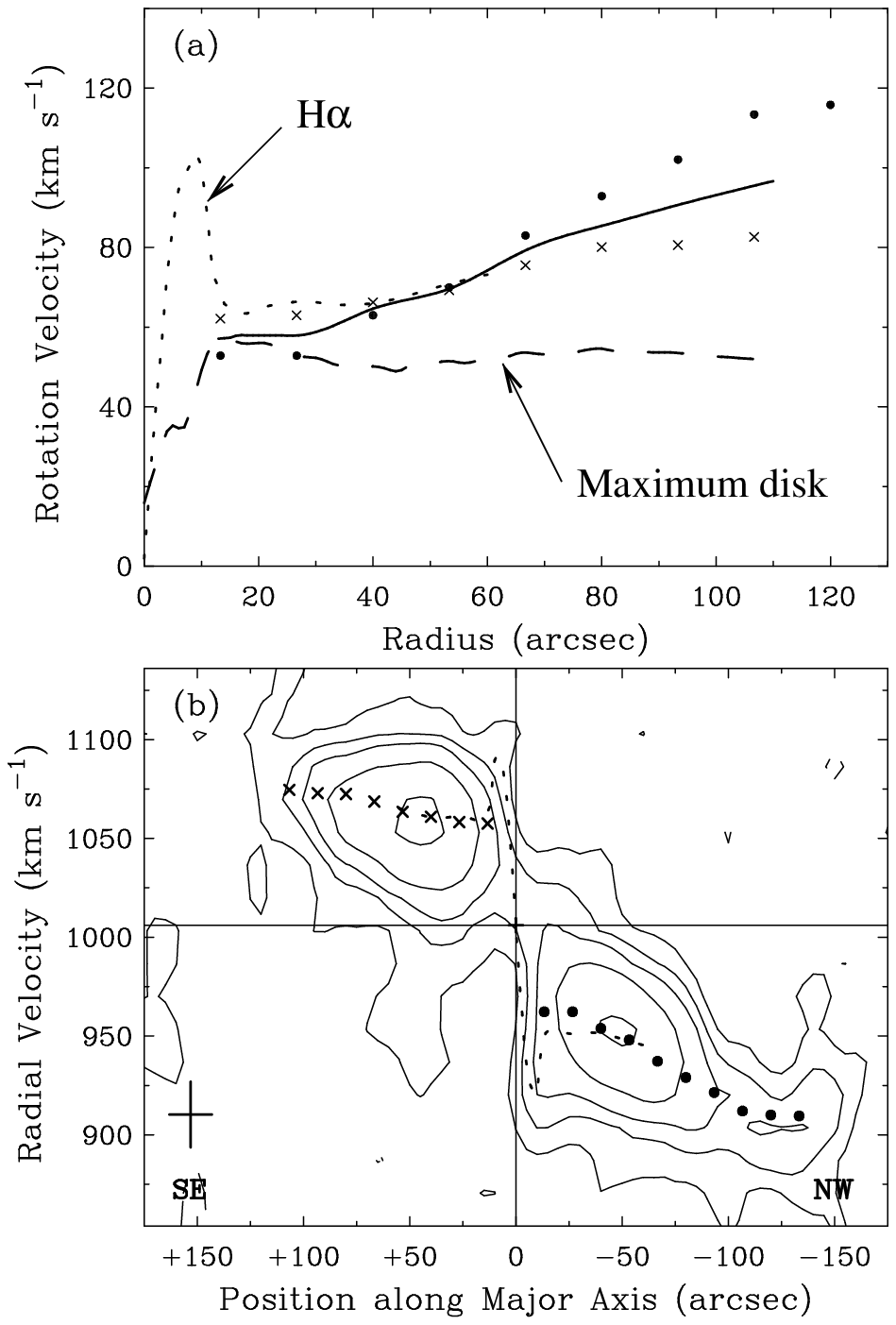}}
	\caption{(a) The solid dots and crosses give the \ion{H}{i}
	rotation velocities of the approaching and receding sides
	respectively, the solid curve shows the mean \ion{H}{i}
	rotation curve. The dashed curve shows the maximum disk (stars
	\& gas) rotation curve, the dotted one is that of the
	H$\alpha$ obtained by van der Kruit (1976), (b) Position-velocity
	map along the major axis (position angle 150 degrees) at 20\arcsec\
	resolution. Contour levels are $-$1.7, 1.7, 3.4, 5.1, 8.5
	and 16.9 mJy beam$^{-1}$. The dots and crosses give the projected
	\ion{H}{i} rotational velocities for the approaching and
	receding sides. The dotted curve is from van der Kruit (1976).}
	\label{fig:rotcur}
\end{figure}

The \ion{H}{i} velocity dispersion is unusually large (ranging from 20
to 40 km s$^{-1}$), especially in the inner high-density regions which
coincide with the stellar disk. \ion{H}{i} position-velocity maps
taken from the 20\arcsec\ data at various position angles (see
e.g. Fig.~\ref{fig:rotcur}b) show this very clearly and reveal the
presence of faint velocity tails even extending to the quadrant of
forbidden velocities (across the systemic velocity line). In the same
areas of the disk, \ion{H}{ii} regions also show large (30--40 km
s$^{-1}$) velocity dispersions and large deviations from circular
motion \citep{vdK76,GS91,MD96}.

A tilted-ring model \citep[cf.][]{B89} was fitted to the velocity
field to determine the center of rotation and the inclination
angle. This led to an inclination of 56$\pm$7 degrees for the
\ion{H}{i} disk, significantly larger than found for the optical disk
(Table~\ref{tab:globalprop}).

The azimuthally--averaged radial density distribution of the \ion{H}{i}
was calculated from the data at 30\arcsec\ resolution by averaging the
signal in circular rings in the plane of the galaxy. The inclination
angle was kept fixed at 56 degrees and the major axis position angle at
150 degrees. The \ion{H}{i} distribution is approximately exponential. The
scalelength ($h_{\ion{H}{i}}$) and the radius at a level of 1
M$_\mathrm{\sun}$ pc$^{-2}$ ($R_{\ion{H}{i}}$) are given in
Table~\ref{tab:derivedprop}. They are similar to those 
found for galaxies of the same Hubble type (Sbc) as \object{NGC 3310}
\citep{V97}.

The tilted-ring model fits were also used to derive the rotational
velocity. Inclination and position angles were kept fixed. The
rotational velocities obtained for an inclination angle of 56 degrees,
a position angle of 150 degrees and systemic velocity of 1006 km
s$^{-1}$ and estimated for each side separately, are presented in
Fig.~\ref{fig:rotcur}a as dots (approaching side) and crosses (receding
side). The \ion{H}{i} rotational velocity is $\sim$60 km s$^{-1}$
at 15\arcsec\ from the center, just outside the optical ring, and
slowly increases further out. Beyond the optical disk, the velocity of
the approaching side rises up to 120 km s$^{-1}$ whereas the
velocities of the receding side level off at 80 km s$^{-1}$. The curve
derived from the H$\alpha$ observations \citep{vdK76} is also shown
(rescaled using the \ion{H}{i} inclination). It shows solid body
rotation in the inner region up to the starburst ring at
$\sim$10\arcsec\, reaching a maximum of $\sim$105 km s$^{-1}$. This is
followed by a sharp drop-off down to 60 km s$^{-1}$ and a slight
increase up to 80 km s$^{-1}$ at 70\arcsec. The sharp drop in the
curve is steeper than Keplerian, indicating that there are
non-circular motions present in the inner region \citep{vdK76}.

\begin{figure}[!t]
	\resizebox{7.5cm}{!}{\includegraphics{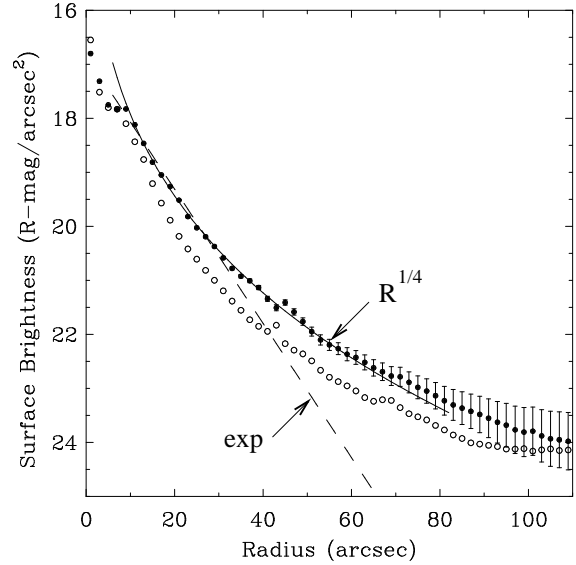}}
	\caption{{\it R}--band surface brightness profile of
	\object{NGC 3310}. The dots are for the inclination and
	position angle of the \ion{H}{i}
	(Table~\ref{tab:derivedprop}), the circles for the inclination
	and position angle of the H$\alpha$
	(Table~\ref{tab:globalprop}).}
	\label{fig:sbprof}
\end{figure}

Figure~\ref{fig:rotcur}b shows the rotation curves of the approaching
and receding sides superposed (after projection to $v_{\mathrm{rot}}$
$\sin i$) on the \ion{H}{i} position-velocity map along the major
axis. The projected H$\alpha$ velocities \citep{vdK76} are also shown
(dotted line). There is good agreement between the \ion{H}{i} and
H$\alpha$: although the \ion{H}{i} data are too coarse to get a good
estimate of the rotation near the center, the lowest contours in the
\ion{H}{i} position-velocity map show that the \ion{H}{i} motion is
consistent with the H$\alpha$ velocities. 
Figure~\ref{fig:rotcur}b shows clearly that, in spite of its overall
rotation, the \ion{H}{i} has a peculiar kinematics characterized by a
large velocity dispersion and asymmetric profiles. The \ion{H}{i}
rotation curve derived here has large uncertainties, especially
because of the large deviations from circular motion. This uncertainty
is of order $20-30$ km s$^{-1}$. However, the agreement of the
\ion{H}{i} and H$\alpha$ curves between 20 and 60 arcsec
(Fig.~\ref{fig:rotcur}) indicates that the amplitude of the projected
rotation curve adopted here is probably correct. The optical and
\ion{H}{i} inclination and position angles
(Tables~\ref{tab:globalprop} and \ref{tab:derivedprop}) are also very
uncertain and do show large discrepancies.

Because of all these problems and uncertainties the standard analysis
with the well-known decomposition in luminous (stars and gas) and dark
components \citep{B89} may be doubtful. In the following, the results
obtained on the mass distribution, the total mass and the
mass-to-light ratio should therefore be taken with caution.

The {\it R}--band surface brightness profile (Fig.~\ref{fig:sbprof}, dots)
was derived from an optical ({\it R}--band) image of \object{NGC 3310}
\citep{SASH01} using the inclination and position angles as determined from the
\ion{H}{i} (Table~\ref{tab:derivedprop}). It is best described by an
{\it R}$^{1/4}$ law (Fig.~\ref{fig:sbprof}, the effective parameters
$\mu_{e}$ and $r_{e}$ are given in Table~\ref{tab:derivedprop}),
except for the inner 10\arcsec\ (the compact nucleus and the starburst
ring). For comparison an exponential disk fit to the bright disk gives
a scalelength $h_{\rm R} = 9\pm1\arcsec$ ($\sim$0.6 kpc) and
$\mu_{0}=16.82\pm0.05$ R--mag arcsec$^{-2}$. The surface brightness
profile as derived using the inclination and position angle of the
optical disk is also shown (circles).

\begin{table}[!h]
	\caption{\ion{H}{i} and optical parameters for \object{NGC 3310}}
\begin{flushleft}
\begin{tabular}{lll}\hline\noalign{\smallskip}\hline\noalign{\smallskip}
\multicolumn{1}{l}{Quantity} & \multicolumn{1}{l}{Units} & \multicolumn{1}{l}{Value} \\ \hline\noalign{\smallskip}
inclination angle (\ion{H}{i})        & degrees                    & $56\pm7$\\
position angle (\ion{H}{i}) (N$\rightarrow$E) & degrees            & $150\pm5$\\
$v_\mathrm{sys}$ (heliocentric)	      & km s$^{-1}$                & $1006\pm4$\\
$M_\mathrm{\ion{H}{i}}^\mathrm{disk}$ & 10$^{9}$ M$_\mathrm{\sun}$ & $2.2\pm0.1$\\
$h_\mathrm{\ion{H}{i}}$	 & arcsec                       & $48\pm4$\\
$R_\mathrm{\ion{H}{i}}$	 & arcsec                       & $109\pm5$\\
$R_{T}^{0}$ mag    		 & mag                          & $10.42\pm0.07$\\
$L_\mathrm{R}$$^{1}$	 & $10^{10}\ {\rm L}_\mathrm{R,\sun}$ & $0.64\pm0.04$\\
$r_{e}$                  & arcsec                       & $12\pm2$\\
$\mu_{e}$                & R--mag arsec$^{-2}$           & $18.22\pm0.05$\\
$M_\mathrm{lum}$ (maximum disk)        & 10$^{10}$ M$_\mathrm{\sun}$ & $0.2$\\
$M_\mathrm{tot}$ (inside r=110\arcsec) & 10$^{10}$ M$_\mathrm{\sun}$ & $2.2$\\
$M_\mathrm{\ion{H}{i}}^\mathrm{disk}$/$L_\mathrm{R}$ &
M$_\mathrm{\sun}$/L$_\mathrm{R, \sun}$ & $0.34\pm0.02$\\
$M_\mathrm{tot}$/$L_\mathrm{R}$ & M$_\mathrm{\sun}$/L$_\mathrm{R, \sun}$ & 3.4\\
$M_\mathrm{\ion{H}{i}}^\mathrm{disk}$/$M_\mathrm{lum}$ & & $1.1$\\
$M_\mathrm{\ion{H}{i}}^\mathrm{disk}$/$M_\mathrm{tot}$ & & $0.1$\\
\hline\noalign{\smallskip}\hline\noalign{\smallskip}
\end{tabular}
\begin{enumerate}
\item The {\it R}--band magnitude is converted to luminosity in
solar units using M$_\mathrm{R,\sun} = 4.31$.
\end{enumerate}
\end{flushleft}
	\label{tab:derivedprop}
\end{table}

The rotation curve for the disk was calculated from the {\it R}--band
photometric profile (Fig.~\ref{fig:sbprof}, dots) assuming a truncated
disk potential \citep{C83} and a mass-to-light ratio constant with
radius. The rotation curve of the gas was derived from the \ion{H}{i}
density distribution after multiplication by a factor of 1.32 to
correct for the presence of helium. The maximal rotation curve of the
luminous matter (gas \& stars) is shown in Fig.~\ref{fig:rotcur}a
(dashed curve). A clear discrepancy with the observed rotation curve
is visible in the outer parts. Such discrepancies are usually taken to
be evidence for the presence of a dark halo. In the inner ring there
is a large difference between the maximal rotation curve and the
H$\alpha$ curve. This would imply a mass discrepancy also
in the inner region and the presence of a dark mass of about
5 10$^{9}$ M$_{\sun}$. However, the sharp drop in the H$\alpha$ curve
is steeper than Keplerian \citep{vdK76}, indicating that at least part
of the discrepancy is caused by non-circular motions.

The mass derived for the maximum disk and the total mass out to
110\arcsec\ (= 7.1 kpc) are given in Table~\ref{tab:derivedprop}. When
compared to regular spiral galaxies of similar rotational velocity
(e.g. Verheijen, 1997) \object{NGC 3310} has a rather large {\it R}--band
luminosity (Table~\ref{tab:derivedprop}) and low values for the
mass-to-light ratio of the maximum disk
($M_\mathrm{lum}$/$L_\mathrm{R}$ = $0.3$ M$_{\sun}$/L$_{\rm R, \sun}$)
and for the global mass-to-light ratio inside 110\arcsec\ 
($M_\mathrm{tot}$/$L_\mathrm{R}$ = 3.4 M$_{\sun}$/L$_{\rm R, \sun}$).
The amplitude of the rotation curve may have been underestimated due
to our choice of inclination angle. This may have led to a factor 2 
underestimate of the maximum disk and total masses.

\section{Discussion \& Conclusions}

The \ion{H}{i} observations presented in the previous section have
provided new evidence bearing on the dynamics of \object{NGC 3310} and
on the origin of the starburst. The main results are the two conspicuous
\ion{H}{i} tails connected to the northern and southern sides of the disk
and the unusually large velocity dispersion of the
\ion{H}{i} in the disk.

The velocity dispersion of the \ion{H}{i} in the disk, of up to
40 km s$^{-1}$, quite large compared to the values of 7 to 12 km
s$^{-1}$ usually found in spiral galaxies, can be attributed for only
a small part to profile broadening caused by differential rotation
inside the instrumental beam. Most of it  must come from an
intrinsically high gas turbulence or, more likely, to deviations from
circular motion, like the streamings seen in H$\alpha$. Such deviations
would cause, because of the relatively large beam (20 arcsec = 1.3 kpc),
broad velocity profiles. The \ion{H}{i} data confirm the results of
optical spectroscopy and indicate that the gaseous disk of \object{NGC
3310} is highly disturbed. Large deviations from circular rotation
must be present all over the disk and it is not clear whether all
non-circular motions are in the plane of the disk.

The two extended tails on opposite sides of the \object{NGC 3310} disk
have similar \ion{H}{i} masses but different morphologies and
kinematics. The one to the south is more extended, reaching out to 51
kpc, broader on the sky and more patchy and has almost constant radial
velocity (close to the systemic velocity). It has lower \ion{H}{i}
surface density and no optical counterpart. Its velocity dispersion is
around 13 km s$^{-1}$. The tail on the northern side is less extended,
half the length, narrower and covers a larger velocity range. Its
velocity dispersion is very similar to that of the southern tail.

The northern tail has been constructed with the assumption that the
\ion{H}{i} emission with anomalous velocities 804 to 854 km s$^{-1}$
on the northern side of \object{NGC 3310} and the \ion{H}{i}
coinciding with the `arrow' form one continuous structure. This is
clearly in contrast with the interpretation of the arrow as a
one-sided jet violently emitted from a compact central source as
proposed by \citet{BS84}. According to the interpretation favoured
here, the chain of compact \ion{H}{ii} regions visible to the north of
the bright optical disk would be part of a curved structure continuing
to the west with the `arrow' and coinciding with the \ion{H}{i}
tail. On Plate 2 of \citet{BS84} the continuity between the arrow and
the chain of northern optical knots seems obvious. This would imply,
however, similar radial velocities of \ion{H}{i} and \ion{H}{ii},
whereas the measurements by \citet{MD96} seem to indicate that the 
\ion{H}{ii} velocities are about 100 km s$^{-1}$ higher. It is
possible that there is confusion with \ion{H}{ii} emission from the
disk. At any rate, it is clear that this point is important for the
interpretation proposed here and should be verified by accurately
determining the optical radial velocities.

In conclusion, it seems that in all their properties the
two \ion{H}{i} tails found associated with NGC 3310 resemble the
tidal tails seen in gravitationally interacting systems and mergers.
It should be noted, however, that in this respect the relative
orientation of the two tails may pose a problem (see below).

All the evidence discussed above --the optical ripples, the disturbed
kinematics of the gaseous disk and the two \ion{H}{i} tails-- seems to
point to a recent merger event. A merger or a major accretion event for
NGC 3310 has been suggested before \citep{BH81,SS88}. One possibility
is the type of encounter in which an Irr I galaxy is being cannibalized
by NGC 3310 as proposed by \citet{BH81}. Any such explanation should,
however, also account for the presence of the two gaseous tails
revealed by the present study. The two equal mass, extended tails, the
\ion{H}{i} gas content and the small total mass seem to point to a
merger between two galaxies of small and comparable masses, of which
at least one gas-rich. Indeed, Schweizer's \citeyearpar{S78} five
characteristics to be expected for a recently merged pair of galaxies
--a pair of long tails, an isolated merger-remnant candidate, a single
nucleus, chaotic motions and the tails moving in opposite directions
-- seem to be all present for NGC 3310. For the fifth condition --the
two tails moving in opposite directions--, there may be a problem
concerning the orientation of the northern tail. It is possible,
however, that the tail is not in one plane and that the optical and HI
pictures are the projection of a more complex structure. The cluster
of northern HII regions does suggest a possible turn of direction from
north-north-east to the south-west \citep[see Plate 2 of Bertola and
Sharp][]{BS84}. The actual merger process may have been more complex
and has, perhaps, involved a third small object as the `bow' and the
other ripples seem to suggest. It is remarkable, at any rate, that
after such an apparently `major' merger event there should still be a
disk. This is clearly different in many respects (in spite of the
remarkable {\it R}$^{1/4}$ photometric profile) from the well studied
merger case \object{NGC 7252}, the `Atoms-for-Peace'
\citep{HGGS94}. The unsettled disk we observe now in \object{NGC 3310}
could then be either a newly formed disk with spiral arms and star
formation going on, or the disturbed disk of one of the progenitors
which has survived the merger and is now undergoing new star
formation. This would argue in favour of the robustness of disks.

The present observations have added a few more pieces to the
interesting puzzle of \object{NGC 3310}. It is clear, however,
that for a better understanding of the origin of its peculiar
morphology and kinematics detailed numerical simulations as done
for example by \citet{HiM95} for \object{NGC 7252} are needed.

\begin{acknowledgements}
We wish to thank Rob Swaters and Marc Balcells for providing the
{\it R}--band image of \object{NGC 3310}, and Tjeerd van Albada and
Piet van der Kruit for helpfull comments. We have used the GIPSY
package, developed at the Kapteyn Astronomical Institute, for the
data reduction and analysis. The Westerbork Synthesis Radio Telescope
is operated by the  Netherlands Foundation for Research in Astronomy
(NFRA / ASTRON), with financial support by the Netherlands Organization
for Scientific Research (N.W.O.).
\end{acknowledgements}

\bibliographystyle{apj}
\bibliography{mymnemonic,article}

\begin{thebibliography}{25}
\expandafter\ifx\csname natexlab\endcsname\relax\def\natexlab#1{#1}\fi

\bibitem[{Balick \& Heckman(1981)}]{BH81}
Balick, B. \& Heckman, T. 1981, A\&A, 96, 271

\bibitem[{Begeman(1989)}]{B89}
Begeman, K.~G. 1989, A\&A, 223, 47

\bibitem[{Bertola \& Sharp(1984)}]{BS84}
Bertola, F. \& Sharp, N.~A. 1984, MNRAS, 207, 47

\bibitem[{Broeils \& van Woerden(1994)}]{BW94}
Broeils, A.~H. \& van Woerden, H. 1994, A\&AS, 107, 129

\bibitem[{Casertano(1983)}]{C83}
Casertano, S. 1983, MNRAS, 203, 735

\bibitem[{de~Vaucouleurs {et~al.}(1991)de~Vaucouleurs, de~Vaucouleurs, Corwin,
  et~al.}]{V91}
de~Vaucouleurs, G., de~Vaucouleurs, A., Corwin, H.~G., et~al. 1991, Third
  Reference Catalogue of Bright Galaxies (New York: Springer)

\bibitem[{Duric {et~al.}(1986)Duric, Seaquist, Crane, \& Davis}]{D86}
Duric, N., Seaquist, E.~R., Crane, P.~C., \& Davis, L.~E. 1986, ApJ, 304, 82

\bibitem[{Grothues \& Schmidt-Kaler(1991)}]{GS91}
Grothues, H.~G. \& Schmidt-Kaler, T. 1991, A\&A, 242, 357

\bibitem[{Hibbard {et~al.}(1994)Hibbard, Guhathakurtha, van Gorkum, \&
  Schweizer}]{HGGS94}
Hibbard, J.~E., Guhathakurtha, P., van Gorkum, J.~H., \& Schweizer, F. 1994,
  AJ, 107, 67

\bibitem[{Hibbard \& Mihos(1995)}]{HiM95}
Hibbard, J.~E. \& Mihos, J.~C. 1995, AJ, 110, 140

\bibitem[{Kikumoto {et~al.}(1993)Kikumoto, Taniguchi, Suzuki, \&
  Tomisaka}]{K93}
Kikumoto, T., Taniguchi, Y., Suzuki, M., \& Tomisaka, K. 1993, AJ, 106, 466

\bibitem[{Lequeux(1971)}]{L71}
Lequeux, J. 1971, A\&A, 15, 30

\bibitem[{Mulder \& van Driel(1996)}]{MD96}
Mulder, P.~S. \& van Driel, W. 1996, A\&A, 309, 403

\bibitem[{Mulder {et~al.}(1995)Mulder, van Driel, \& Braine}]{MDB95}
Mulder, P.~S., van Driel, W., \& Braine, J. 1995, A\&A, 300, 687

\bibitem[{Pastoriza {et~al.}(1993)Pastoriza, Dottori, Terlevich, Terlevich, \&
  Diaz}]{P93}
Pastoriza, M.~G., Dottori, H.~A., Terlevich, E., Terlevich, R., \& Diaz, A.~I.
  1993, MNRAS, 260, 177

\bibitem[{Rhee \& van Albada(1996)}]{RA96}
Rhee, M.-H. \& van Albada, T.~S. 1996, A\&AS, 115, 407

\bibitem[{Schweizer(1978)}]{S78}
Schweizer, F. 1978, in Structure and Properties of Nearby Galaxies, ed.
  E.~Berkhuijsen \& R.~Wielebinski (Reidel), pp.279

\bibitem[{Schweizer \& Seitzer(1988)}]{SS88}
Schweizer, F. \& Seitzer, P. 1988, ApJ, 328, 88

\bibitem[{Smith {et~al.}(1996)Smith, Neff, Bothun, et~al.}]{S96}
Smith, D.~A., Neff, S.~G., Bothun, G.~D., et~al. 1996, ApJ, 473, L21

\bibitem[{Swaters {et~al.}(2001)Swaters, van Albada, Sancisi, \& van~der
  Hulst}]{SASH01}
Swaters, R.~A., van Albada, T.~S., Sancisi, R., \& van~der Hulst, J.~M. 2001,
  A\&A, submitted

\bibitem[{Telesco \& Gatley(1984)}]{TG84}
Telesco, C.~M. \& Gatley, I. 1984, ApJ, 284, 557

\bibitem[{van~der Kruit(1976)}]{vdK76}
van~der Kruit, P.~C. 1976, A\&A, 49, 161

\bibitem[{van~der Kruit \& de~Bruyn(1976)}]{vdKB76}
van~der Kruit, P.~C. \& de~Bruyn, A.~G. 1976, A\&A, 48, 373

\bibitem[{Verheijen(1997)}]{V97}
Verheijen, M. A.~W. 1997, PhD thesis, University of Groningen

\bibitem[{Walker \& Chincarini(1967)}]{WC67}
Walker, M.~F. \& Chincarini, G. 1967, ApJ, 147, 416

\end{thebibliography}

\end{document}